\documentclass[aps,prl,reprint,amsmath,amssymb,nofootinbib,superscriptaddress]{revtex4-2}

\usepackage{graphicx}
\usepackage{booktabs}

\graphicspath{{fig/}}

\newcommand{\nubar}{\bar\nu}
\newcommand{\GeV}{\,\mathrm{GeV}}
\newcommand{\MeV}{\,\mathrm{MeV}}
\newcommand{\fm}{\,\mathrm{fm}}
\newcommand{\DS}{\Delta S}
\newcommand{\Msun}{M_\odot}
\newcommand{\rhoz}{\rho_0}
\newcommand{\Mmax}{M_{\max}}

\begin{document}

\title{Accelerator neutrinos as a probe of in-medium hyperon potentials}

\author{J.\,A.\ Nowak}
\thanks{ORCID: \href{https://orcid.org/0000-0001-8637-5433}{0000-0001-8637-5433}}
\email[Corresponding author: ]{j.nowak@lancaster.ac.uk}
\affiliation{School of Physics and Astronomy, Lancaster University,
Lancaster LA1 4YB, United Kingdom}

\date{\today}

\begin{abstract}
Charged-current (anti)neutrino interactions create $\Lambda$ and $\Sigma$ hyperons
inside the nucleus, so the hyperons that reach a detector have crossed nuclear matter
and carry the imprint of the in-medium potentials $U_Y(\rho)$ that also govern the
appearance of hyperons in neutron stars. In the StrangeMC simulation of argon, the
reconstructed $\Lambda$ yield and momentum and a kaon-vetoed $\Sigma^+$ tag respond to
$U_\Lambda$ and $U_\Sigma$ at SBND and DUNE energies. A detector-level
forecast, with the production cross sections, fluxes and efficiencies as nuisance
parameters and the muon kinematics calibrating the production model, determines
$U_\Lambda$ at $\rho^*\simeq0.82\rhoz$, independently of the low-density exponent of
the potential, to $1.5\MeV$ statistically and $\pm31\MeV$ in total. The largest item,
$18\MeV$, is the convention by which the cascade applies the potential. The
measurement is a check, with different systematics, of the hypernuclear constraint,
$\pm3.3\MeV$ at $\rho^*$, for momentum-independent potentials; $U_\Sigma$ is not
determined. In the relativistic mean-field equations of state GM1 and FSU2H, anchored to the
hypernuclear and $\Sigma$-atom depths, the maximum neutron-star mass depends on the
$\Lambda NN$ repulsion, which these data do not constrain.
\end{abstract}

\maketitle

\textit{Introduction.}---Pulsar timing has established neutron stars of about
$2\,\Msun$ (PSR~J1614$-$2230~\cite{demorest2010}, J0348$+$0432~\cite{antoniadis2013},
J0740$+$6620~\cite{fonseca2021}), which require a stiff equation of state (EOS) at
several times the saturation density $\rhoz=0.16\fm^{-3}$. Above about $2\rhoz$ it
becomes favourable to convert neutrons into hyperons, which relieves the Fermi pressure,
softens the EOS and generically lowers the maximum mass $\Mmax$ below the observed
values, the hyperon puzzle~\cite{chatterjee2016,tolos2020,burgio2021}. The outcome
depends on the in-medium single-particle potential $U_Y(\rho)$ of each hyperon $Y$: its
depth at saturation sets where hyperons appear, and its density dependence above
saturation, dominated by three-body $YNN$ forces, sets how much they soften the
matter. At saturation the $\Lambda$ is bound, $U_\Lambda(\rhoz)\approx-28\MeV$ from
$\Lambda$ hypernuclei~\cite{gal2016}, and the $\Sigma$ is repelled,
$U_\Sigma(\rhoz)\approx+30\MeV$ from $\Sigma^-$ atoms and $(\pi^-,K^+)$
spectra~\cite{saha2004}; heavy-ion $\Lambda$ flow probes higher
densities~\cite{ohnishi2022,nara2022,Jinno:2025lfu}.

Accelerator (anti)neutrinos give a further handle at and below saturation. A
charged-current interaction creates the hyperon inside the nucleus, through a weak
vertex on a bound nucleon, and the hyperon reaches the detector after crossing nuclear
matter. Hyperon production off nuclei with final-state interactions (FSI) has been
computed for antineutrino quasi-elastic production on oxygen with a cascade and a
vertex potential~\cite{Sobczyk:2019uej} and off nuclei with a $\Lambda$-nucleus
potential and $\Sigma\!\to\!\Lambda$ conversion~\cite{benitez2024}, and MicroBooNE has
measured antineutrino quasi-elastic $\Lambda$ production on argon~\cite{microboone2023lambda}.
The present calculation varies the potential in a simulation of argon and carries it to
a detector-level forecast with production-model, hyperon--nucleon (YN), transport,
nuclear initial-state and detector uncertainties.

\textit{The neutrino handle.}---In antineutrino-mode beams (reverse horn current,
RHC) the hyperons come mainly from the Cabibbo-suppressed ($\DS=1$) quasi-elastic
reaction $\nubar_\mu N\to\mu^+Y$, $Y\in\{\Lambda,\Sigma^0,\Sigma^-\}$, which supplies
$77\%$ of the strange final states at SBND and $49\%$ at DUNE, where associated $YK$
production supplies $45\%$. Charged-current deep-inelastic strange production, not
included in these shares or in the forecast templates, would take $60\%$ of the DUNE
sample and leave the quasi-elastic reaction $66\%$ at SBND and $18\%$ at DUNE, so a DUNE
analysis has to model or reject it. The strangeness-changing two-body current
$\nubar_\mu NN\to\mu^+YN$~\cite{Nowak:2026twobody} adds $17\%$ (SBND) and $21\%$
(DUNE) to the quasi-elastic hyperon yield. Neutrino-mode beams (forward horn current,
FHC) produce hyperons mainly with a kaon. The potential acts on the escaping hyperons in four
ways: (i)~it shifts their energy on exit, an attractive $\Lambda$ well lowering and a
repulsive $\Sigma$ well raising the momentum; (ii)~it captures slow $\Lambda$'s and
moves the escaping spectrum relative to the detector thresholds; (iii)~it shifts the
thresholds of $\Sigma N\leftrightarrow\Lambda N$ conversion; and (iv)~it changes the
rate of a kaon-less $\Sigma^+$. The weak vertex on a single nucleon cannot make a
$\Sigma^+$ ($\Delta S=\Delta Q$), so in quasi-elastic events a $\Sigma^+$ comes from
FSI, in $0.80\%$ (SBND RHC) and $1.06\%$ (DUNE RHC) of the hyperon events. Its
competitor, secondary $\pi N\to\Sigma^+K$, carries a kaon: the signal-to-background
ratio of $0.55$ and $0.68$ before cuts rises to $1.8$ and $2.2$ with a charged-kaon
veto and to $3.6$ and $4.4$ with a $K^0_S$ veto as well. The two-body current also
makes a $\Sigma^+$ at the weak vertex on a proton pair~\cite{Nowak:2026twobody}.

\textit{Simulation.}---StrangeMC~\cite{companionPRD} simulates strange-particle
production and transport in (anti)neutrino--$^{40}$Ar interactions for the SBND and
DUNE beams. The production channels are calibrated to published cross
sections~\cite{singh2006,alam2010,alam2013,fatima2025} and compared with
MicroBooNE~\cite{microboone2025}; the nuclear initial state and the intranuclear
cascade come from the LUNAR package~\cite{Nowak:2026czc}. The hyperon potential is
\begin{equation}
  U_Y(\rho)=U_Y(\rhoz)\,(\rho/\rhoz)^\gamma
            + c_Y\big[(\rho/\rhoz)^\beta-(\rho/\rhoz)^\gamma\big] ,
  \label{eq:UY}
\end{equation}
with $\beta=2$. At low density $U_Y\simeq t_{YN}\rho$, with $t_{YN}$ fixed by the
forward $YN$ amplitude~\cite{Friedman:2007zza}; $\gamma=1$ is that limit and
$\gamma\neq1$ its leading correction. The $\rho^2$ term is the leading three-body
contribution, which chiral effective field theory and quantum Monte Carlo calculations
find dominant at $2$--$3\rhoz$~\cite{Lonardoni:2014bwa,Haidenbauer:2016vfq,Gerstung:2020ktv};
for $\gamma=1$ Eq.~\eqref{eq:UY} is the linear-plus-quadratic form of hypernuclear
fits~\cite{Millener:1988hp}, written so that $U_Y(\rhoz)$ remains an explicit parameter.
The term with $c_Y$ vanishes at $\rhoz$ but not below it: at $0.82\rhoz$ and $\gamma=1$ it
shifts $U_\Lambda$ by $-0.15\,c_\Lambda$.

The potential is applied in the hyperon cascade, which performs elastic $YN$
scattering, Pauli-blocked knockout and the charge-conserving conversion channels. On
escape the energy is shifted by the potential at the production density $\rho_v$,
$E_{\rm out}=E_{\rm in}+U_Y(\rho_v)$, and a hyperon with $E_{\rm out}\le m_Y$ is
captured; two alternatives are gradient-force transport and the convention of the
NuWro cascade~\cite{golan2012,thorpe2021}, in which $U_Y(\rho_v)$ is subtracted at the start of
the cascade and restored on exit, so that a hyperon that does not rescatter leaves with
$E_{\rm in}$. At the vertex the struck nucleon is off shell, with energy
$M_N-e_{\rm rem}$, and the hyperon is produced on shell, so at fixed momentum transfer
the removal energy is carried by the muon and $U_Y$ acts only in the cascade. The
nucleon self-energy reaches the hyperon only through the momentum distribution of the
struck nucleon and the sampling of the energy transfer, so a change of $e_{\rm rem}$
does not mimic a change of $U_Y$ one for one. Capture is a transport-level proxy for hypernucleus formation; the
captured fraction is not fitted.

\begin{figure}[t]
\centering
\includegraphics[width=\columnwidth]{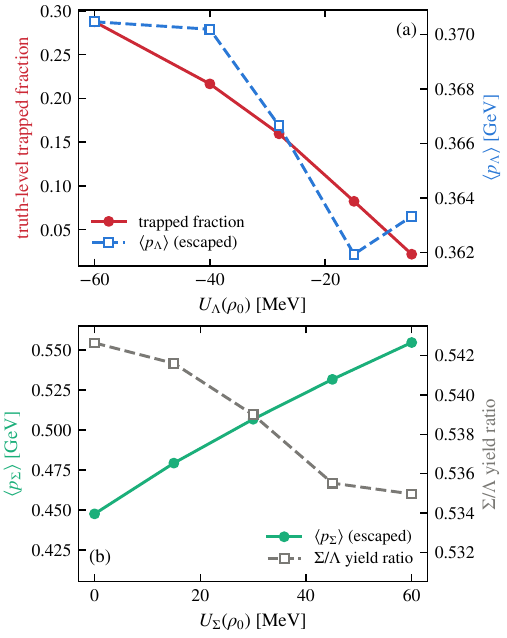}
\caption{Response of the SBND RHC hyperon observables to the potential. (a)~The
captured $\Lambda$ fraction (truth level) and the mean momentum of the escaping
$\Lambda$'s versus $U_\Lambda(\rhoz)$. (b)~The mean escaping-$\Sigma$ momentum and the
$\Sigma/\Lambda$ ratio versus $U_\Sigma(\rhoz)$. Each point has $1.2\times10^4$
hyperons; the statistical uncertainty is $0.002$--$0.005$ on the fraction and
$0.003\GeV$ on the momenta.}
\label{fig:potsens}
\end{figure}

\textit{Sensitivity.}---Around $U_\Lambda(\rhoz)=-28\MeV$, $U_\Sigma(\rhoz)=+30\MeV$
and $\gamma=1$ (Fig.~\ref{fig:potsens}), deepening $U_\Lambda$ at SBND RHC from $-5$ to
$-60\MeV$ raises the captured fraction monotonically from $0.02$ to $0.29$ and the mean
momentum of the escaping $\Lambda$'s from $0.363$ to $0.370\GeV$ (between $-5$ and
$-15\MeV$ the change is within the $0.003\GeV$ statistical uncertainty), because slow
$\Lambda$'s are preferentially captured; raising $U_\Sigma$ from $0$ to $+60\MeV$ raises the mean escaping-$\Sigma$
momentum from $0.447$ to $0.555\GeV$ at a nearly constant $\Sigma/\Lambda$ ratio.

\textit{Forecast.}---Each escaped hyperon is decayed and passed through a schematic
liquid-argon response ($5\%$ momentum resolution, $21$ and $10\MeV$ proton and pion
thresholds, $90\%$ track efficiency, a $\Lambda\to p\pi^-$ tag). The observables are
the reconstructed-$\Lambda$ fraction $f_\Lambda$, its mean momentum
$\langle p_\Lambda\rangle$ and the $\Sigma^+$ fraction, on a grid of
$(U_\Lambda,U_\Sigma,\gamma)$ for the four beams. The number of produced hyperons is
known only as well as the cross sections, fluxes and efficiencies, which enter as
nuisance parameters ($20\%$ on the quasi-elastic and $50\%$ on the associated cross
section, $7.5\%$ on each flux, $3\%$ on the $\Lambda$ and $30\%$ on the $\Sigma^+$
efficiency), and the two-body yield enters with a $\pm30\%$ normalisation. The yields
then carry little information; it comes mainly from $\langle p_\Lambda\rangle$, whose
spectrum the exit potential moves relative to the thresholds, and from the
$\Sigma^+/\Lambda$ ratio. The production-model parameters (quasi-elastic form factors,
associated channel mix, composition of the DUNE samples) are profiled, with the mean
momentum, angle and $Q^2$ of the muon in the $\Lambda$-tagged events constraining them.

Hyperons are produced throughout the nucleus, with mean production density
$0.66\rhoz$, so $U_Y(\rhoz)$ and $\gamma$ enter through a near-degenerate combination.
The data determine $U_\Lambda$ at the decorrelation density $\rho^*=0.82\rhoz$, where
the fitted value does not depend on $\gamma$; the six density dependences tested shift
it relative to the true value by at most $1.4\MeV$, and the Millener--Dover--Gal form by
$0.9$--$1.9\MeV$ with the production model profiled; $U_\Lambda(\rhoz)$ follows only with an assumption on $\gamma$.
The statistical precision is $1.5\MeV$ (Table~\ref{tab:budget});
profiling the production model widens it to $10.0\MeV$, and the systematic items bring
the total to $\pm31\MeV$. The largest is the transport convention: applying the
potential from the start of the cascade, as NuWro does, moves $U_\Lambda(\rho^*)$ by
$+18\MeV$. The YN items reflect a $\Sigma^-p\to\Lambda n$ cross section in the cascade
that is half of the J-PARC E40 data~\cite{miwa2022} and a low-momentum
$\Lambda N$ elastic cross section below the data. Replacing the initial state (argon
spectral function, local Fermi gas, constant separation energy,
$e_{\rm rem}\pm5\MeV$) moves $U_\Lambda(\rho^*)$ by $8\MeV$: extra binding goes mainly
into the muon, and at fixed centre-of-mass angle the escaping $\Lambda$ energy rises by
$0.2$--$0.4\MeV$ per MeV of removal energy, where a one-to-one degeneracy with $U_\Lambda$ would lower it by
$1\MeV$. Halving the YN and production uncertainties gives about $28\MeV$. With exact YN
input, production priors four times tighter and a $1\%$ momentum scale, the transport
convention and the initial state leave about $21\MeV$, and about $12\MeV$ with the convention
settled as well. The hypernuclear
constraint at the same density is $\pm3.3\MeV$, so the neutrino measurement is a check
of $U_\Lambda$ below saturation, through a different reaction and momentum range
($0.4$--$1.0\GeV/c$); its independence is not complete, since the YN cross sections also
enter the hypernuclear and $\Sigma$-atom extractions. The total assumes momentum-independent
potentials. The hypernuclear value refers to zero momentum, and an effective mass
$m^*/m=0.9$ ($0.8$) in $U_\Lambda$ at $\rhoz$ would move $U_\Lambda(\rho^*)$ by $-27$ ($-60$)$\MeV$
in the fit with the hyperon observables alone, so the comparison needs a model of the momentum dependence.
The forecast is also signal-only and uses representative beam energies.

\begin{table}[t]
\caption{One-standard-deviation uncertainty on $U_\Lambda(\rho^*)$,
$\rho^*=0.82\rhoz$, and on $U_\Sigma(\rhoz)$ at fixed $\gamma$, in MeV, for the four
beams combined with the production model profiled and the muon kinematics
constraining it. Group entries are quadrature sums; the total adds the profiled width,
the groups and the curvature of the profiled fit in quadrature. The statistics row has
the two-body normalisation fixed.}
\label{tab:budget}
\begin{ruledtabular}
\begin{tabular}{lrr}
item & $U_\Lambda(\rho^*)$ & $U_\Sigma(\rhoz)$ \\
\hline
statistics                                   & $1.5$  & $5.8$ \\
production model profiled                    & $10.0$ & $15.9$ \\
YN cross sections                            & $17.5$ & $86$ \\
transport (NuWro convention $18.0$)          & $18.0$ & $55$ \\
nuclear initial state                        & $8.0$  & $11$ \\
detector (momentum scale $11.0$)             & $11.1$ & $33$ \\
two-body kinematics, density shape, other    & $1.7$  & $36$ \\
curvature                                    & $6.9$  & $20$ \\
\hline
total                                        & $31$   & $117$ \\
\end{tabular}
\end{ruledtabular}
\end{table}

$U_\Sigma$ is not determined: its statistical precision, $6\MeV$, is overwhelmed by the
YN and transport items, for a total of $108$--$117\MeV$ against the $\pm20\MeV$ of
$\Sigma$-atom analyses. In the cascade $U_\Sigma$ is a real, momentum-independent
potential and the $\Sigma$ absorption is carried by the conversion cross sections, so
with present YN input the neutrino data do not separate the two.

\begin{figure}[t]
\centering
\includegraphics[width=\columnwidth]{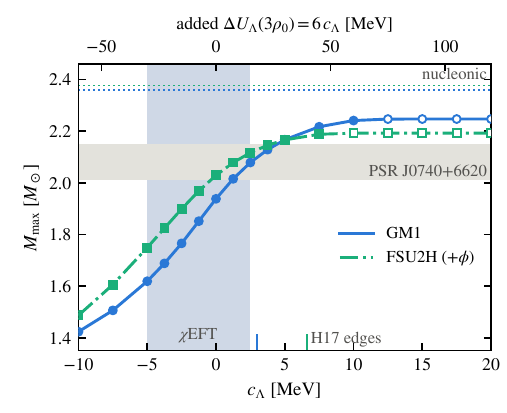}
\caption{Maximum mass of $np\Lambda\Sigma^-$ stars versus the three-body strength
$c_\Lambda$ of Eq.~\eqref{eq:UY}, with the hyperon couplings anchored to
$U_\Lambda(\rhoz)=-28\MeV$ and $U_\Sigma(\rhoz)=+30\MeV$, for GM1 and FSU2H with the
$\phi$ meson; the upper axis gives the added $U_\Lambda$ at $3\rhoz$. The band marks the
chiral-EFT range with $\Lambda NN$ forces, the horizontal band the mass of
PSR~J0740$+$6620, and the dotted lines the nucleonic maximum masses.}
\label{fig:mmax}
\end{figure}

\textit{Neutron-star maximum mass.}---The same depths enter nonlinear
$\sigma$--$\omega$--$\rho$ relativistic mean-field EOSs for $n,p,\Lambda,\Sigma^-$ with
electrons and muons in $\beta$ equilibrium and an SLy crust: GM1~\cite{glendenning1991},
a standard reference point of the hyperon-puzzle literature~\cite{Weissenborn:2011kb}, which
is stiff (incompressibility $301\MeV$, symmetry-energy slope $94\MeV$), and
FSU2H~\cite{Tolos:2016hhl,Tolos:2017lgv} with the $\phi$ meson ($238$ and $44.5\MeV$).
The hyperon vector couplings follow SU(6), and the scalar coupling is fixed by requiring
the symmetric-matter potential at $\rhoz$ to equal the depth,
$U_Y(\rhoz)=-g_{\sigma Y}\sigma(\rhoz)+g_{\omega Y}\omega_0(\rhoz)$, so only
$U_Y(\rhoz)$ enters, through $g_{\sigma Y}$; $g_{\omega Y}$ and $\gamma$ do not. The
three-body term enters as $c_Y[(n_B/\rhoz)^2-n_B/\rhoz]$ in the hyperon energies. At
the hypernuclear and $\Sigma$-atom depths and $c_\Lambda=0$, $\Mmax=1.94\,\Msun$ (GM1)
and $2.03\,\Msun$ (FSU2H) for $np\Lambda\Sigma^-$ matter. A three-body strength within
the chiral-EFT band with $\Lambda NN$ forces~\cite{Gerstung:2020ktv},
$-5\le c_\Lambda\le2.5\MeV$, gives $1.62$--$2.08$ and $1.75$--$2.12\,\Msun$
(Fig.~\ref{fig:mmax}), and a joint fit with a uniform prior over that range gives $\Mmax=1.83$
($1.66$--$2.00$) and $1.96$ ($1.80$--$2.07$)~$\Msun$ at $68\%$. The neutrino data do not move these
values: at fixed $c_\Lambda$ the potentials at saturation spread $\Mmax$ by
$0.02$--$0.05\,\Msun$, an order of magnitude below the spread between EOS models, and
whether these EOSs support $2\,\Msun$ depends on the $\Lambda NN$ repulsion. The radius
and tidal deformability of a $1.4\,\Msun$ star, $14.0$\,km and $895$ for GM1 and $13.3$\,km
and $763$ for FSU2H, against the GW170817 bound
$\Lambda_{1.4}\lesssim580$--$720$~\cite{ligo2017gw170817,abbott2018eos,abbott2019properties},
are properties of the nucleonic EOS.

\textit{Conclusion.}---Hyperons produced by accelerator neutrinos inside argon test the
$\Lambda$ potential below saturation, at $\rho^*\simeq0.82\rhoz$, through a weak
production vertex and a momentum range different from those of hypernuclear
spectroscopy. With the production model profiled, the projected SBND and DUNE data reach
$\pm31\MeV$, limited mainly by the transport convention and the YN cross sections;
with the convention settled, exact YN input, a tighter production model and a $1\%$
momentum scale the total would fall to about $12\MeV$, before any momentum dependence of
the potential. $U_\Sigma$ and the high-density potential are not determined,
and the neutron-star maximum mass follows from the EOS model and the $\Lambda NN$
repulsion. The simulation, the full systematic study and the EOS calculations are given
in Ref.~\cite{companionPRD}.

\begin{acknowledgments}
Although the author is a member of the DUNE and SBND collaborations, all views
presented here are his and not those of the collaborations as a whole. This work was
supported by the Science and Technology Facilities Council (STFC) Lancaster EPP
Consolidated Grant 2025--2029: UKRI2846.
\end{acknowledgments}

\paragraph{Data availability.} The StrangeMC simulation and the analysis and EOS
scripts underlying this work are available from the author on reasonable request.

\bibliographystyle{apsrev4-2}
\bibliography{refs}

\end{document}